\documentclass[conference]{IEEEtran}
\IEEEoverridecommandlockouts
\usepackage{cite}
\usepackage{amsmath,amssymb,amsfonts}
\usepackage{algorithmic}
\usepackage{graphicx}
\usepackage{textcomp}
\usepackage{xcolor}
\usepackage{siunitx}
\usepackage{subfigure}
\usepackage[numbers,sort&compress]{natbib}
\usepackage{booktabs}
\usepackage{hyperref}

\def\BibTeX{{\rm B\kern-.05em{\sc i\kern-.025em b}\kern-.08em
    T\kern-.1667em\lower.7ex\hbox{E}\kern-.125emX}}

\begin{document}

\title{Passive Respiration Detection via mmWave Communication Signal Under Interference\\
}





\author
{\IEEEauthorblockN{Kehan Wu\IEEEauthorrefmark{1}\IEEEauthorrefmark{2}\thanks{\IEEEauthorrefmark{1}Both authors contributed equally to this work.}, Renqi Chen\IEEEauthorrefmark{1}\IEEEauthorrefmark{2}, Haiyu Wang\IEEEauthorrefmark{2}, Chenqing Ji\IEEEauthorrefmark{2}, Jiayuan Zhu\IEEEauthorrefmark{2} and Guang Wu\IEEEauthorrefmark{2}\IEEEauthorrefmark{3}}
\IEEEauthorblockA{\textit{\IEEEauthorrefmark{2}Department of Electronic and Electrical Engineering, College of Engineering, Southern University} \\
\textit{of Science and Technology, Shenzhen, 518055, China} \\
\IEEEauthorrefmark{3}Corresponding authors. e-mails:  \href{wug@sustech.edu.cn}{wug@sustech.edu.cn}
}
}

\maketitle

\begin{abstract}
Recent research has highlighted the detection of human respiration rate using commodity WiFi devices. Nevertheless, these devices encounter challenges in accurately discerning human respiration amidst the prevailing human motion interference encountered in daily life. To tackle this predicament, this paper introduces a passive sensing and communication system designed specifically for respiration detection in the presence of robust human motion interference. Operating within the 60.48 GHz band, the proposed system aims to detect human respiration even when confronted with substantial human motion interference within close proximity.
Subsequently, a neural network is trained using the collected data by us to enable human respiration detection. The experimental results demonstrate a consistently high accuracy rate over 90\% of the human respiration detection under interference, given an adequate sensing duration. Finally, an empirical model is derived analytically to achieve the respiratory rate counting in 10 seconds.
\end{abstract}

\section{Introduction} 

The ability to accurately and reliably detect respiratory patterns has long been a crucial aspect of healthcare monitoring, sleep analysis, and physical activity tracking. A person's respiration provides valuable insights into his overall well-being.  For example, a respiratory rate value greater than 27 breaths per minute (bpm) as been identified as a significant predictor of cardiac arrest and is utilized in the prediction of conditions such as pneumonia or lower respiratory tract infections \cite{nicoloImportanceRespiratoryRate2020,ruminskiAnalysisParametersRespiration2016,chourpiliadisPhysiologyRespiratoryRate2023}. Traditionally, respiration detection has relied on the use of sensors or wearable devices, which often impose constraints on the individual's freedom of movement and may lead to discomfort or inconvenience \cite{massaroniContactBasedMethodsMeasuring2019}.

Passive sensing, a branch of Integrated Sensing and Communication (ISAC), has paved the way for device-free respiration monitoring (DFRM) \cite{vanegasSensingSystemsRespiration2020, wangIntegratedSensingCommunication2022a}. This innovative approach eliminates the need for physical contact by capturing and processing signals influenced by subtle movements associated with respiration. Importantly, passive sensing has demonstrated its ability to accurately capture the motion of the chest caused by respiration without  significantly occupying bandwidth.

There have been several existing methods to estimate respiration rate using non-contact techniques, with most of them relying on channel state information (CSI) or received signal strength (RSS) \cite{nakamuraWiFiBasedFallDetection2022,wangResilientRespirationRate2020,zengFullBreatheFullHuman2018,zengMultiSenseEnablingMultiperson2020}. However, the CSI methods typically utilize only a small part of the signal, resulting in increased delay and compromised accuracy. Moreover, CSI method has to cancel the frequency offset between the transmitter and receiver, realizing the synchronization between them.

Currently, most DFRM systems employ commercial WiFi \cite{nakamuraWiFiBasedFallDetection2022,wangResilientRespirationRate2020,zengFullBreatheFullHuman2018,zengMultiSenseEnablingMultiperson2020}, offering convenience and ease of implementation. However, the detection precision of these systems is limited by the wavelength with the following restriction:
\begin{equation}
f_{d}={\frac{v c o s(\theta)}{\lambda}}
\end{equation}
where $f_{d}$ is Doppler Frequency, $\theta$ is the angle between the direction of velocity and the direction of arrival of the signal, $\lambda$ is the wavelength of the signal. 

To overcome the limitations associated with lower sensitivity, the utilization of millimeter waves offers distinct advantages in DFRM systems. Millimeter waves have shorter wavelengths, enabling higher precision in detecting subtle respiratory movements. Furthermore, millimeter-wave technology is extensively employed in next-generation communication systems, creating an opportunity for synergistic integration.

In general, there are two types of interference:
\begin{enumerate}
    \item Static Environmental Interference: Environmental interference encompasses external factors that potentially compromise the accuracy and reliability of a measurement or detection process, such as the presence of reflected signals from the static environment. Within the domain of respiration detection, interference stemming from the static environment can significantly impact the precision of measurements.
    \item Motion Interference: The interference caused by human motion is
induced through random human motion around the detected target
in the area during respiration detection. Movements such as walking, running, or even slight body motions can introduce artifacts and distort the respiration signal. Motion interference poses a challenge because it can lead to inaccurate measurements and make it difficult to extract the true respiratory pattern.

\end{enumerate} 

To address these challenges, we propose a DFRM system based on passive millimeter radar with a strong anti-interference ability, acceptable processing speed, and improved accuracy. It will offer a powerful solution for advancing both healthcare monitoring and communication capabilities.

The subsequent sections of this paper are structured as follows. Section \ref{Overview of the System} presents an introduction to the mmWave passive sensing and communication system. The signal processing algorithm for passive sensing and the methodology for approximating classification accuracy are expounded in Section \ref{System Model}. Section \ref{Experiment} outlines the system implementation, accompanied by the demonstration and discussion of experimental results. Lastly, the paper is concluded in Section \ref{Conclusion}.

\section{Overview of the System}
\label{Overview of the System}
\begin{figure}[!htbp]
  \centering
  \includegraphics[width=0.45\textwidth]{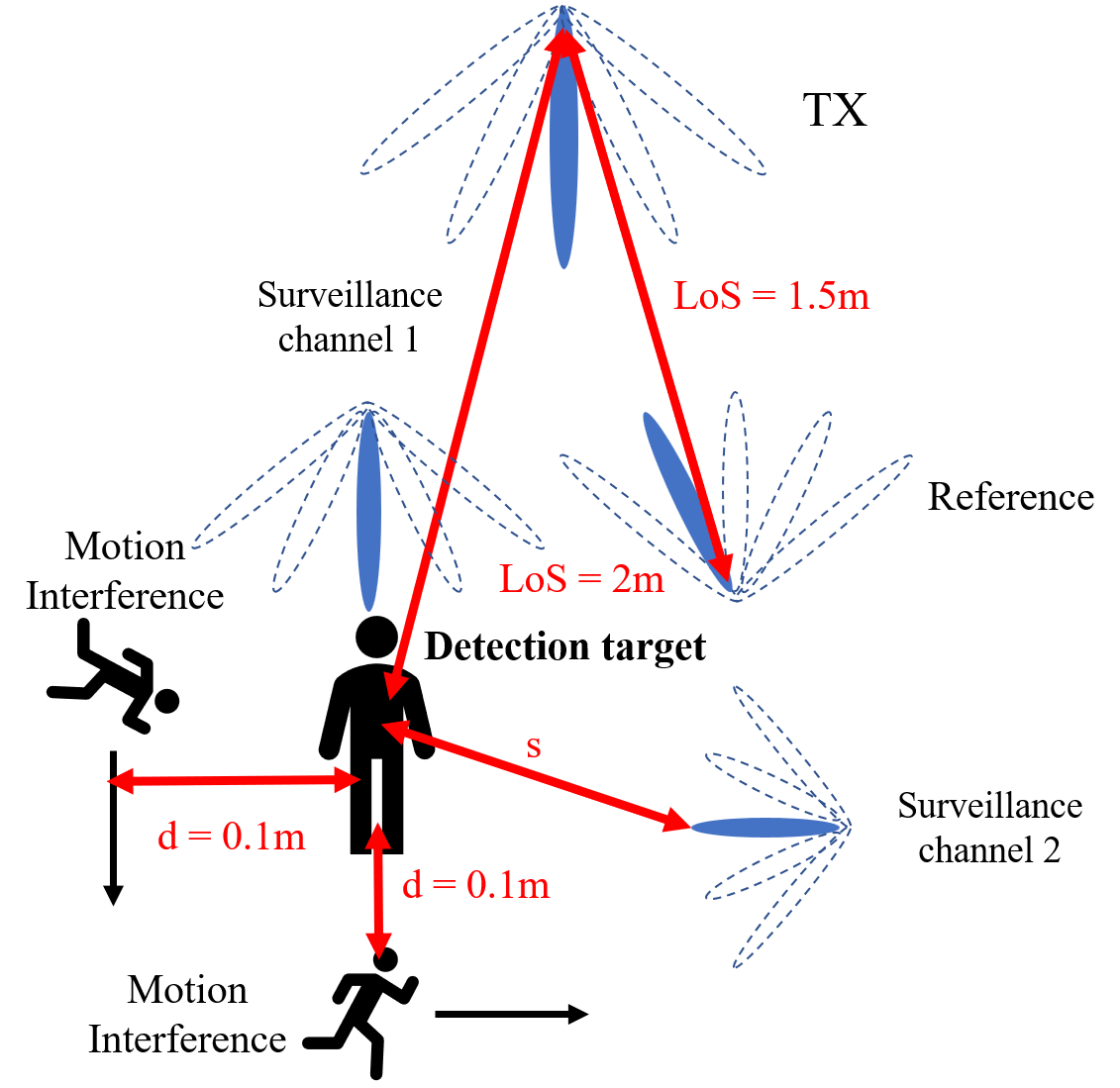}
  \vspace{-4mm}
  \caption{Model of the Sensing System.}
  \label{fig:System Model}
\end{figure}
A model of passive sensing and communication system working on the mmWave waveband is elaborated in Fig. \ref{fig:System Model}.  
The transmitter portion transmit the communication signal with a carrier frequency of 60.48 GHz over the downlink. 

The transmitter part is composed of a Sivers millimeter wave phased array block with 16 antennas and a USRP baseband processing block.
The receiver portion is consists of 3 different channels. 
To be specific, the first channel demonstrated as the reference channel is settled to receive the communication signal, 
the second and third channel denoted as surveillance channel. Both are used to receive the reflected signal from the chest. 
The transmitter delivers the 
communication signal to the receiver without Doppler frequency in the reference channel, and the receiver towards the reference channel receive 
signal to obtain the transmission signal without interference. The surveillance channels receive the echo signal from the moving target in the environment,
and the reflected signal is processed by the receiver to obtain the Doppler information of the target \cite{wang_multi_2018}.
The surveillance channels are used to detect the Doppler frequency in the surroundings, and the Doppler frequency can derive the approximate human respiration rate.
And s represents the distance between the target and the closest receiver which is a variable in different experiment.

\begin{figure}[!htbp]
  \centering
  \includegraphics[width=0.3\textwidth]{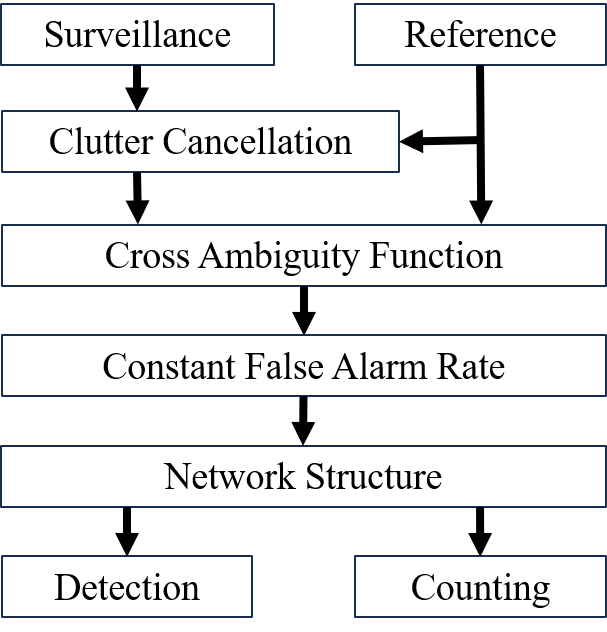}
  \vspace{-3mm}
  \caption{Signal Processing Diagram.}
  \label{fig:System Diagram}
\end{figure}

Subsequently, a clutter cancellation method is applied to mitigate the internal interference within the signals obtained from both the surveillance channels and the reference channel. The processed signals from these channels are then subjected to the Cross Ambiguity Function (CAF) and Constant False Alarm Rate (CFAR) algorithms to extract the Doppler frequency information. These resulting Doppler frequency values are subsequently fed into a neural network that has been trained using the collected data, enabling the detection of human respiration.

The system's primary objective is twofold: firstly, to detect the presence of human respiration in the surrounding environment, and secondly, to accurately count the respiratory rate.

\section{System Model}
\label{System Model}
\subsection{Signal Model}
The transmitter transmits the communication signal $s(t)$ to the receiver, the received signal via the reference channel, denoted as $y_{r}(t)$, can be expressed in (\ref{eq1})
\begin{equation}
y_{r}(t)=\alpha_{r}s(t-\tau_{r})e^{-j2\pi f_{\varDelta} t}+n_{r}(t),\quad0\leq t\leq\mathrm{T}, 
 \label {eq1}
\end{equation}
where $\alpha_{r}$ and $\tau_{r}$ denote the attenuation coefficient and time delay of the reference channel respectively, $f_{\varDelta}$ denotes the frequency offset between the transmitter and receiver, $n_{r}(t)$ denotes the noise and other interference (e.g., sensing target, the surrounding static scattering clusters), $\mathrm{T}$ denotes the duration of the transmit signal. Based on the scattering loss, the interference power is often much weaker than the signal power in the reference channel.


The received signal via the $i$-th $(i = 1, 2)$ surveillance channel, denoted as $y_{s,i}(t)$, can be expressed in (\ref{eq2})
\begin{equation}
\begin{aligned}
y_{s,i}(t)=& \alpha_{s,i}^{\mathrm{tar}}(t)s\left(t-\tau_{s,i}^{\mathrm{tar}}(t)\right)e^{-j2\pi f_{\varDelta} t}e^{-j2\pi f_{s,i}^{\mathrm{tar}}(t)t}  \\
&+\sum_{k=1}^{L_s}\alpha_{s,i}^ks(t-\tau_{s,i}^k)e^{-j2\pi f_{\varDelta} t}+n_{s,i}(t),\quad0\leq t\leq\mathrm{T},
\end{aligned}
\label{eq2}
\end{equation}
where $\alpha_{s,i}^{\mathrm{tar}}(t)$, $\tau_{s,i}^{\mathrm{tar}}(t)$, and $f_{s,i}^{\mathrm{tar}}(t)$ denote the time-varying attenuation coefficient, time delay and Doppler frequency of the scattered path off the sensing target respectively, $f_{\varDelta}$ denotes the frequency offset between the transmitter and receiver, $L_s$ denotes the number of paths from clutter interference based on static scattering, $\alpha_{s,i}^k$ and $\tau_{s,i}^k$ denote the attenuation coefficient and time delay of the $k$-th path of static clutter interference respectively, and $n_{s,i}(t)$ denotes the noise.


\subsection{Signal Processing of Passive Sensing}
The received signals from reference and surveillance channels
are sampled at the baseband with a period $\mathrm{T_s}$, which can be
expressed by
\begin{equation}
    y_{r}[n]=y_{r}(n\mathrm{T_s})\text{ and }y_{s,i}[n]=y_{s,i}(n\mathrm{T_s}),
\end{equation}
where $n=1,2,...,\mathrm{T}/\mathrm{T_s}$, $i=1,2$ which denote the two surveillance channels.
The received signals in two surveillance channels undergo the clutter cancellation firstly to mitigate the clutter interference originating from the static environment. Subsequently, it is processed by the Cross Ambiguity Function and Constant False Alarm Rate algorithms to calculate the Doppler Frequency within the vicinity. The resulting information is then forwarded to a network architecture to facilitate both the detection and counting tasks of human respiration.
\subsubsection{Adaptive Clutter Cancellation Algorithm}
Note that the signal components with zero Doppler frequency in $y_{s,i}[n]$ can seriously interfere the estimation of the target Doppler frequency $f_{s,i}^{\mathrm{tar}}(t)$ in (\ref{eq2}).
To solve this problem, the Adaptive Clutter Cancellation Algorithm is used to eliminate the interference of reflected signal from the 
static environment \cite{sun_through-wall_2021}. 
The algorithm encompass the Least Mean Square (LMS)
, recursive minimum algorithm, and normalized least mean square algorithm. In this work, 
an adaptive direct path clutter cancellation algorithm based on LMS is used.

The $\mathbf{V}_r$ is the matrix of the delayed reference signal
 in the reference channel $y_{r}[n]$ expressed by
\begin{equation}
    \mathbf{V}_r=\left[ \begin{matrix}
        y_r[1]&		y_r[0]&				\cdots&		y_r[-P+2]\\
        y_r[2]&		y_r[1]&				\cdots&		y_r[-P+3]\\
        \vdots&		\vdots&				\vdots&		\vdots\\
        y_r[N]&		y_r[N-1]&		 		\cdots&		y_r\left[ N-P+1 \right]\\
    \end{matrix} \right],
    \label{eq3}
\end{equation}
where $P$ is the number of delays, $N$ is the number of sample points in one Coherent Integration
Time (CIT).

Suppose $y_{s,i}$ is the received signal vector of the surveillance channel in one CIT time, then $\mathbf{V}_{s,i}$ is the matrix of
the $i$-th surveillance channel represented by
\begin{equation}
    \mathbf{V}_{s,i}=[y_{s,i}[1],y_{s,i}[2],\cdots ,y_{s,i}[N]]^T
    \label{eq4}
\end{equation}

According to the LMS method, the weight matrix can be represented by

\begin{equation}
    \mathbf{K}=\left( \mathbf{V}_{r}^{H}\mathbf{V}_{r} \right) ^{-1}\mathbf{V}_{r}^{H}\mathbf{V}_{s,i}
    \label{eq5}
\end{equation}

The received signal of the surveillance channel in one CIT after eliminating clutter interference is denoted as $y_{s,i}^{*}$, which can be expressed by

\begin{equation}
    y_{s,i}^{*}= y_{s,i}-\mathbf{V}_{r}\mathbf{K}.
    \label{eq6}
\end{equation}

\subsubsection{Doppler Frequency Estimation Algorithm}
CAF is a crucial radar signal processing tool that plays a vital 
role in representing the matched filter response. It provides valuable information about the time 
delay presented in the echo signal as well as the energy of the Doppler-shifted echo component. 
By estimating the correlation between signals collected from the reference channel and surveillance channel, the CAF enables the 
estimation of target parameters such as time delay and Doppler frequency shift. The expression 
of the CAF in one CIT is shown below:
\begin{equation}
    R_{i}\left( f_d \right) =\max_{\tau} \sum_{n=0}^{N-1}{y_{s,i}^{*}}\left[ n \right]  y_r^{\dagger}\left[ n-\tau \right]e^{-j2\pi f_dnT_s},
      \label{eq7}
\end{equation}
where $i$ denotes the $i$-th surveillance channel, $\left(.\right)^{\dagger}$ denotes the complex conjugate, $f_d$ denotes the Doppler frequency, $N$ denotes the number of sample points in one CIT, and $\tau$ denotes the matched time delay.

By examining the CAF, it becomes apparent that 
the CAF traverses various combinations of time delay and Doppler frequency 
within their respective feasible ranges. When the values of $\tau$ and $f_d$ align with the actual time delay and Doppler 
frequency shift between the surveillance channel and the reference channel, a 
distinct local peak marked as $R_i\left( f_d \right)$ emerges.
Utilizing this information, one can estimate the 
precise values of the time delay between 
the reference channel and surveillance channel and the Doppler frequency of the sensing target. However, in this work since we only care the feature extraction of the Doppler frequency, the time delay $\tau$ is not considered as a parameter in (\ref{eq7}).

\subsection{Methodologies to Mitigate Interference}
Due to factors such as scattering from different body parts and noise interference in the environment, the Doppler frequency is not unique.
The approach to cancel the possible errors used in this work is the two-dimensional CFAR algorithm (2D-CA-CFAR) \cite{kronauge_fast_2013}. In the 2D-CA-CFAR algorithm, the unit that 
needs to be compared with the threshold is referred to as the cell under 
test (CUT). Typically, an equal number of units in the surrounding of the CUT 
are used to compute the background noise power, and these units are called 
training units ($w$). In addition, there are several units between the CUT 
and the training units that do not participate in the calculation of the 
background noise power, and these units are referred to as guard units. 
The threshold $\beta=\alpha P_n$ is determined by the background noise power $P_n$ and 
the threshold factor $\alpha$. The background noise power is the mean of the 
training units and can be written as
\begin{equation}
    P_n=\frac{1}{N}\sum_{w\in W}w,
\end{equation}
where $N$ is the number of training units, and $W$ is the set of training 
units.The threshold factor $\alpha$ is determined by the pre-defined false alarm 
rate and can be expressed as
\begin{equation}
    \alpha=N(P_{fa}^{-1/N}-1),
    \end{equation}
where $P_{fa}$ represents the false alarm rate.

The CFAR algorithm dynamically adjusts the 
threshold for target detection based on the level of noise present in 
the environment. This adaptive adjustment helps to reduce the chances of 
false detection, missed detection, and the false alarms. Furthermore, 
the CFAR algorithm effectively handles the scenarios where no Doppler 
frequency is present, ensuring accurate detection and improved performance 
overall.

\subsection{Respiration Detection Based on Deep Learning}
For the two primary research tasks of respiration detection, namely the relationship between the detection of respiration presence or absence under interference and sensing duration, and the counting of the respiratory rate under interference, deep learning methods are chosen for the analysis. For the consideration of generalizability and time cost, ResNet \cite{he2016deep} is adopted in this work.

The basic building block of ResNet is residual block, which consists of a skip connection and batch normalization. The skip connection directly adds the input to the output, enabling the network to transfer the gradient well to avoid the problem of gradient disappearing and gradient explosion during the training process.

\begin{figure}[!htbp]
  \centering
  \includegraphics[width=0.5\textwidth]{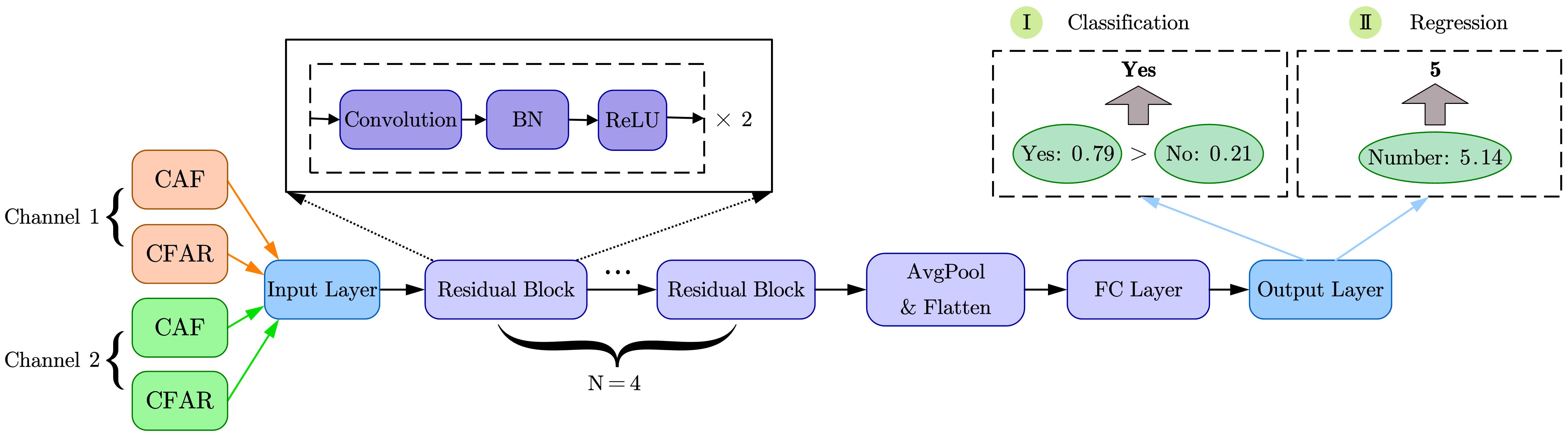}
  \vspace*{-6mm}
  \caption{Network Structure.}
  \label{fig:resnet}
\end{figure}

The network structure is elaborated in Fig. \ref{fig:resnet}, which uses four residual blocks. The input of the ResNet are Time-Doppler spectrograms obtained from CAF and CFAR results of two surveillance channels as raw CAF results provide more micro information and CFAR provide more macro information. The output of the ResNet is different for respective tasks. For the detection of respiration presence or absence within sensing duration, treat it as a binary classification task and then the output layer has two nodes. On the other hand, for the respiratory rate counting problem, treat it as a regression task and the output layer has only one node. Since we only consider the case where the number of breaths is an integer, the output of the model will be rounded. Then we can evaluate the accuracy of the regression results which contain different discrete counting values as a classification problem.

\section{Experiment}
\label{Experiment}
\subsection{Implementation}
\begin{figure}[!htbp]
  \centering
  \includegraphics[width=0.45\textwidth]{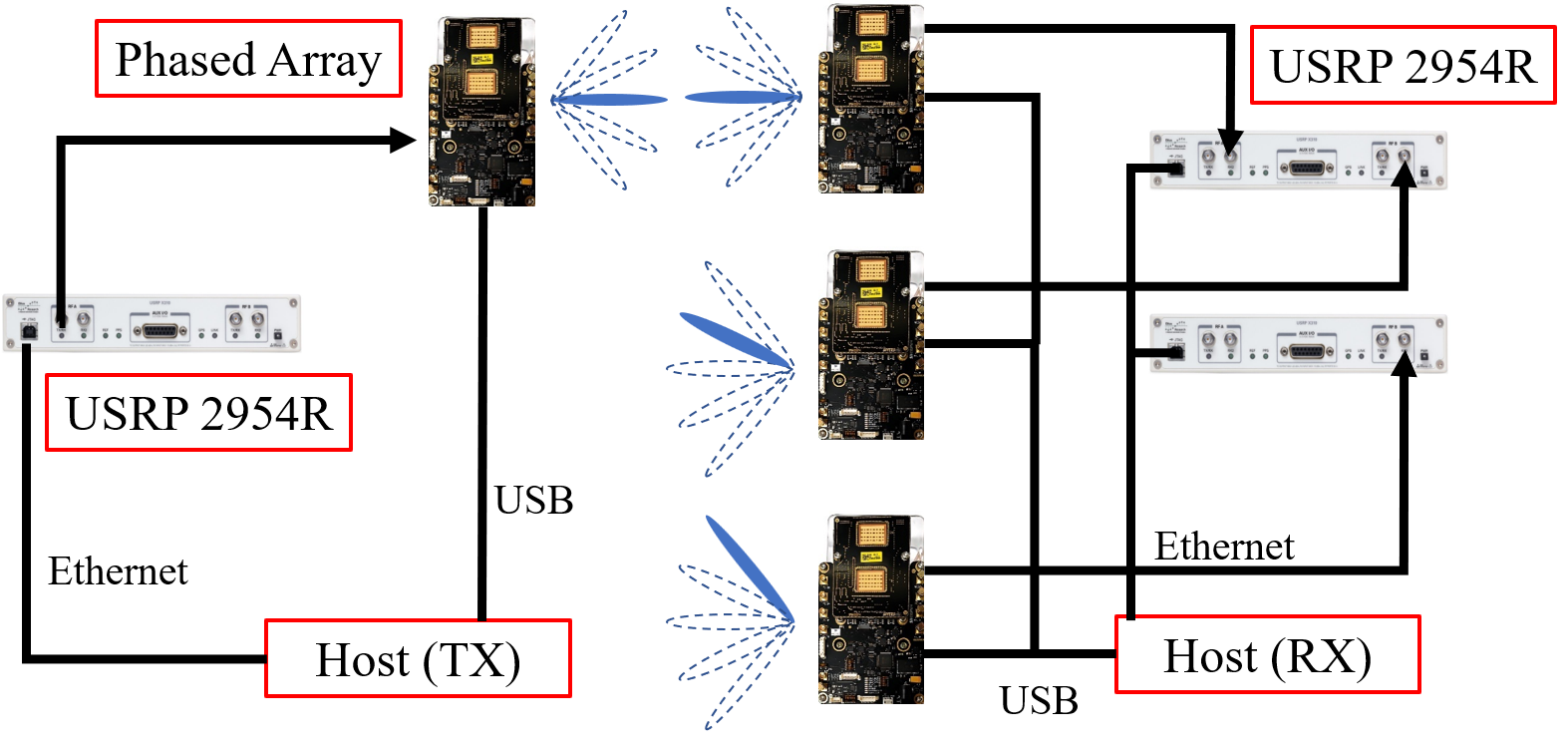}
  \vspace{-5mm}
  \caption{Block diagram of system implementation.}
  \label{fig:Block diagram of system implementation.}
\end{figure}
The system diagram, as depicted in Fig. \ref{fig:Block diagram of system implementation.}, presents an overview of the 
overall system configuration. In the transmitter section, a NI 
USRP-2954R 
is utilized to generate an intermediate frequency (IF) 
signal centered at 500 MHz. This IF signal is subsequently upconverted 
to 60.48 GHz and transmitted using a Sivers 60.48 GHz phased array. The transmission signal comprises a training sequence 
lasting 16 $\mu s$ and is succeeded by an Orthogonal Frequency Division Multiplexing (OFDM)-modulated data payload with a duration of 200 $\mu s$. 
On the receiver side, three 60 GHz phased arrays called Ref, Sur1 and Sur2 are connected to three channels of two USRPs, enabling the reception of signals from both the reference channel 
and two surveillance channels. Note that Sur2 outputs the clock signal to Sur1 and Ref which synchronize the clocks and eliminate the carrier frequency offset (CFO) among them and these three phased arrays are controlled by laptops, 
facilitating collaborative beam switching functionality. After a predefined duration, the receiver host read the data samples collected by two USRPs through Gigabit Ethernet port. In this work, the duration of CIT is 0.1 s.

\begin{figure}[!htbp]
  \centering
  \includegraphics[width=0.48\textwidth]{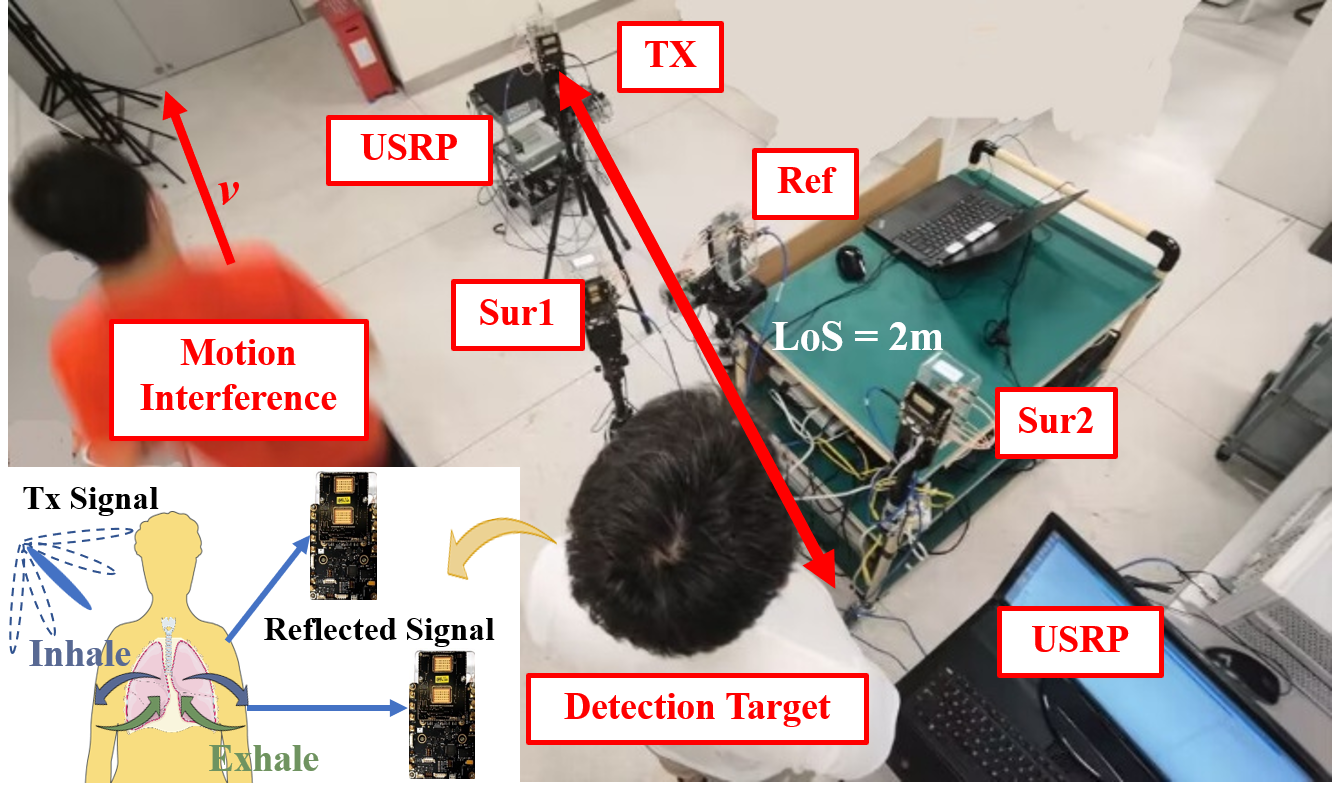}
  \vspace{-8mm}
  \caption{Experiment Layout.}
  \label{fig:ex_lay}
\end{figure}

Fig. \ref{fig:ex_lay} depicts the experimental layout, which corresponds to the model presented in Fig. \ref{fig:System Model},  wherein the Line-of-Sight (LoS) channel is utilized for the purpose of respiration detection. The distance from the transmitter to the detection target (experimenter 1) is 2 m. And experimenter 2 moving around in the vicinity serves as interference. The distance between the target and the closest receiver is changing in different experimental design. 

\subsection{Dataset}
\begin{table*}[]
\centering
\caption{Datasets setup}
\label{table1}
\begin{tabular}{|cccccccl|}
\hline
\multicolumn{8}{|c|}{Detection Dataset}                                                                                                                                                                                                                     \\ \hline
\multicolumn{2}{|c|}{Positive Samples}                                    & \multicolumn{2}{c|}{Negative Samples}                                     & \multicolumn{2}{c|}{Train:Test}                                       & \multicolumn{2}{c|}{K-Fold Method} \\ \hline
\multicolumn{2}{|c|}{200}                                                 & \multicolumn{2}{c|}{200}                                                  & \multicolumn{2}{c|}{8:2}                                              & \multicolumn{2}{c|}{5-Fold} \\ \hline
\multicolumn{8}{|c|}{Counting Dataset}                                                                                                                                                                                                                      \\ \hline
\multicolumn{1}{|c|}{2 Respirations} & \multicolumn{1}{c|}{3Respirations} & \multicolumn{1}{c|}{4 Respirations} & \multicolumn{1}{c|}{5 Respirations} & \multicolumn{1}{c|}{6 Respirations} & \multicolumn{1}{c|}{Train:Test} & \multicolumn{2}{c|}{K-Fold Method} \\ \hline
\multicolumn{1}{|c|}{200}            & \multicolumn{1}{c|}{200}           & \multicolumn{1}{c|}{200}            & \multicolumn{1}{c|}{200}            & \multicolumn{1}{c|}{200}            & \multicolumn{1}{c|}{8:2}        & \multicolumn{2}{c|}{5-Fold} \\ \hline
\end{tabular}
\end{table*}

The experiment involves sensing two types of human respiration: 
detection and counting the human respiration under interference. 
The interference caused by human motion is induced through random 
walking around the detected target in the surrounding area. Noticed that there are no restrictions on the respiratory intensity of the target, i. e. both deep and calm breathing are allowed. In experimental scenario, the detection target can breath freely. To ensure the robustness of respiratory detection system, different volunteers (variations in height, gender, body shape and so forth) are invited as the detection target. Furthermore, a group of volunteers is instructed to walk around the target with varying directions and speeds. 

Initially, we detect the presence of human respiration amidst the interference, considering periods of 2.5, 5, 7, and 10 seconds. Subsequently, we quantify the occurrence of human respiration under the influence of interference, conducting multiple counts ranging from 2 to 6 instances. Each situation of the experiment is sampled 100 times with the passive sensing system in LoS scenarios, resulting in two datasets, illustrated in Table \ref{table1}.

\subsection{CAF and CFAR Results}
Fig. 6(a) and Fig. 6(e) illustrate the time-Doppler spectrograms and CFAR results for the detection of human respiration in the presence of motion interference. Conversely, Fig. 6(b) and Fig. 6(f) display the time-Doppler spectrograms and CFAR results for human respiration in the presence of static environment interference. In contrast, Fig. 6(c)(d)(g)(h) solely depict the spectrograms and CFAR results of interference signals. These spectrograms provide clear visual representations of distinct patterns that distinguish human respiration across various time periods and interference, thereby emphasizing noticeable disparities between interference signals and respiration signals.

\begin{figure}[ht] 
	\centering  
	\vspace{-0.35cm} 
	\subfigtopskip=2pt 
	\subfigbottomskip=2pt 
	\subfigcapskip=-5pt 
	\subfigure[]{
		\label{3}
		\includegraphics[width=0.48\linewidth]{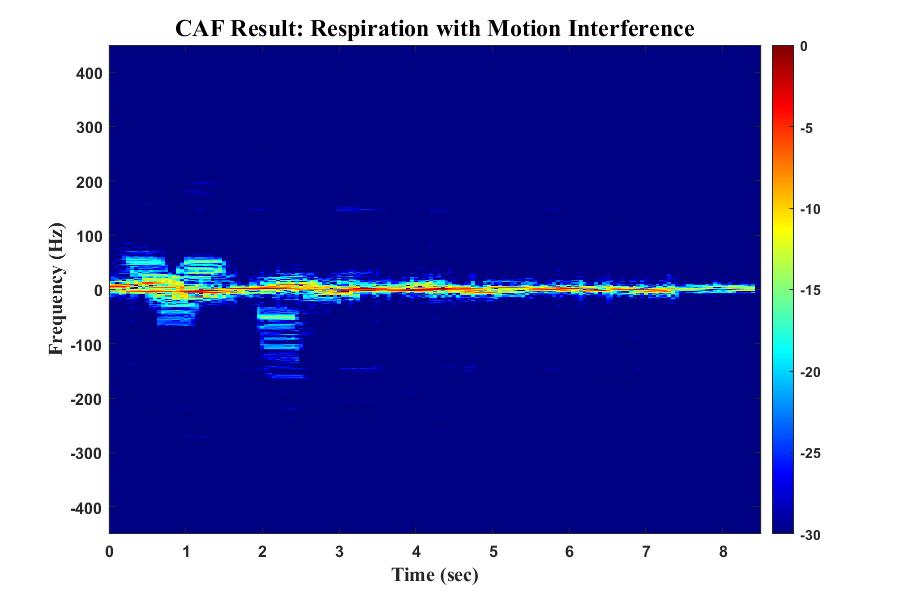}}
    \hspace*{-2mm}
    \subfigure[]{
		\label{7}
		\includegraphics[width=0.48\linewidth]{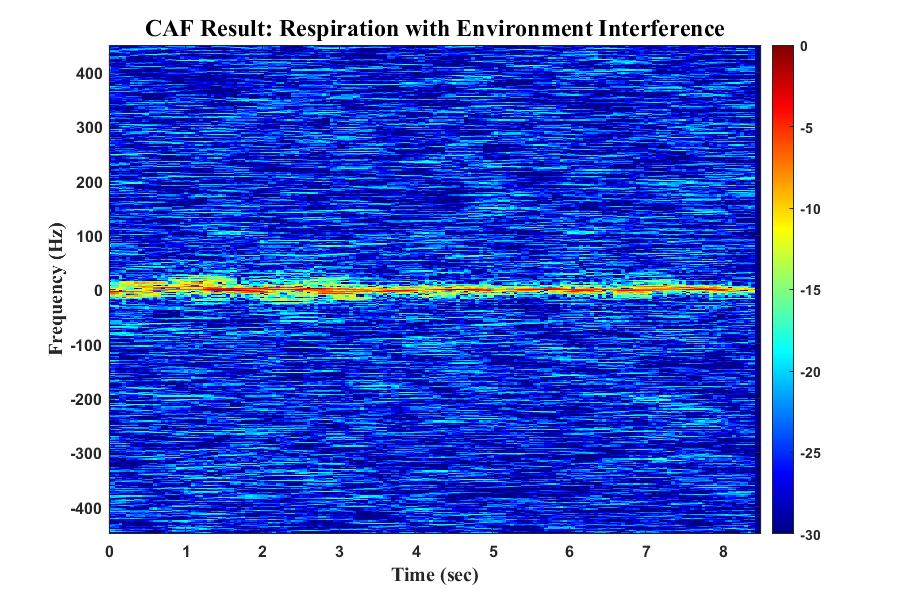}}
    \hspace*{-2mm}
    	\subfigure[]{
		\label{star}
		\includegraphics[width=0.48\linewidth]{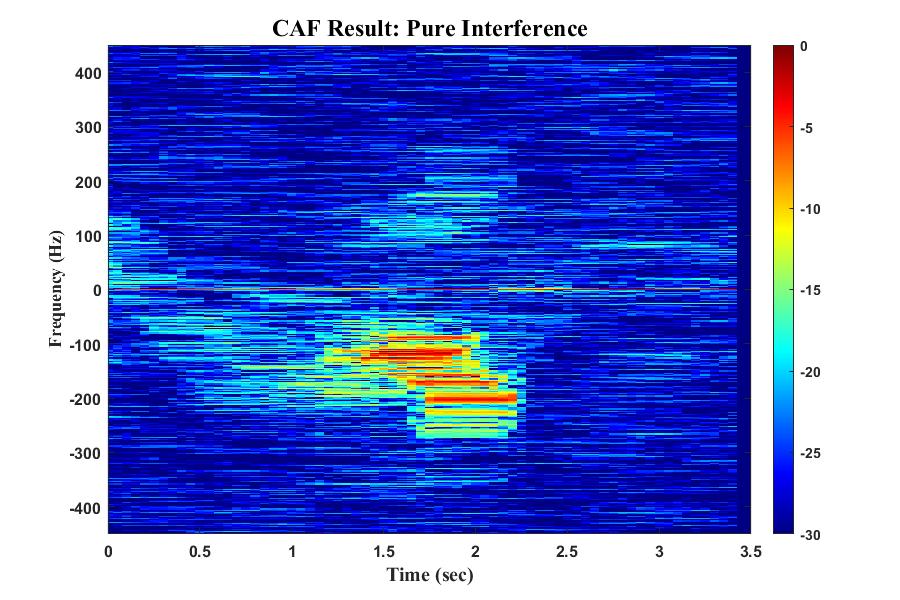}}
    \hspace*{-2mm}
	\subfigure[]{
		\label{star}
		\includegraphics[width=0.48\linewidth]{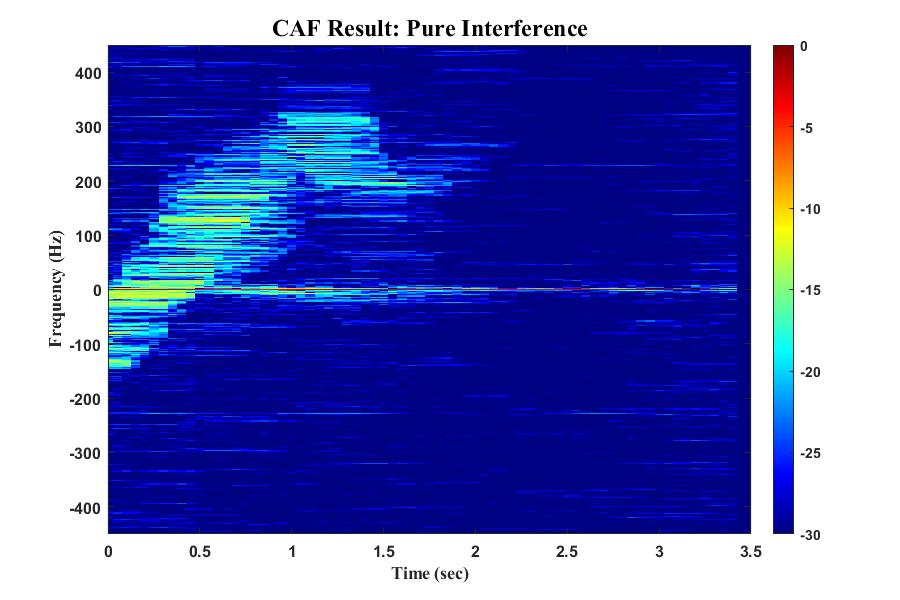}}
    \hspace*{-2mm}
    	\subfigure[]{
		\label{3}
		\includegraphics[width=0.48\linewidth]{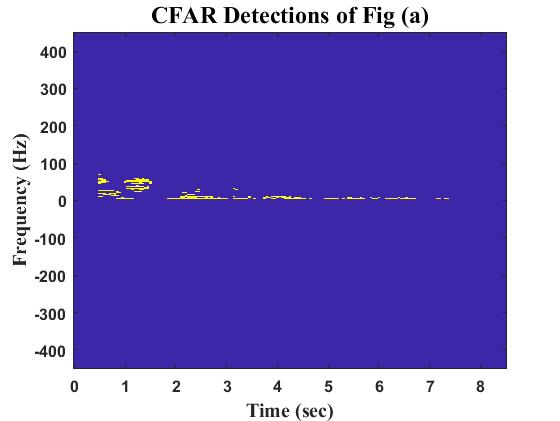}}
    \hspace*{-2mm}
    \subfigure[]{
		\label{7}
		\includegraphics[width=0.48\linewidth]{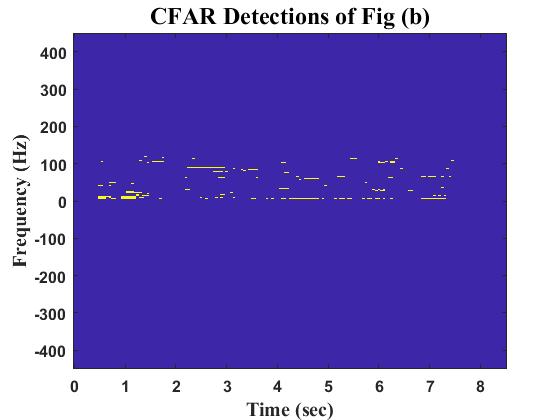}}
    \hspace*{-2mm}
    	\subfigure[]{
		\label{star}
		\includegraphics[width=0.48\linewidth]{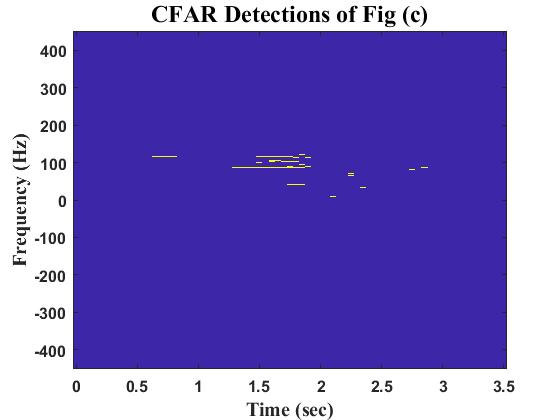}}
    \hspace*{-2mm}
	\subfigure[]{
		\label{star}
		\includegraphics[width=0.48\linewidth]{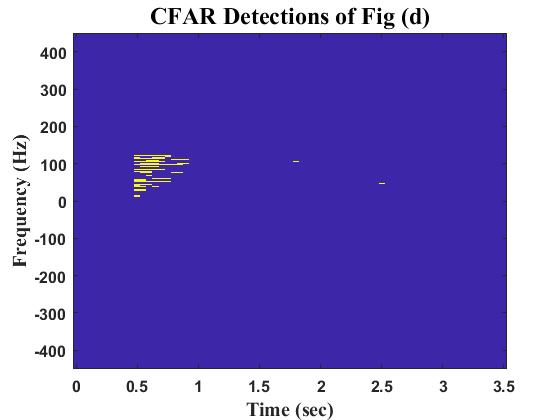}}
	\caption{
Fig (a) and (e) represent the results of Cross Ambiguity Function (CAF) and Constant False Alarm Rate (CFAR) algorithms applied to respiration signals in the presence of human motion interference. Fig (b) and (f) display the CAF and CFAR results of respiration signals affected by static environment interference. On the other hand, Fig (c), (d), (g), and (h) exhibit the CAF and CFAR results obtained from environments where no respiration activity is present.}
	\label{result}
\end{figure}

\subsection{Respiration Detection Results}
\subsubsection{Detection}
\begin{figure}
  \centering
  \includegraphics[width=0.4\textwidth]{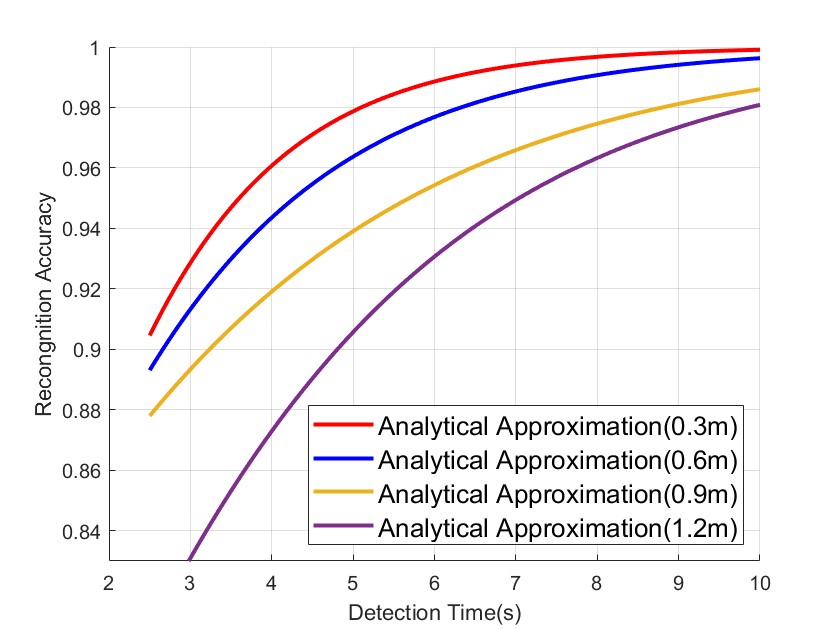}
  \vspace{-3mm}
  \caption{Classification accuracy versus the sensing time.}
  \label{fig:det_time}
\end{figure}

This work investigates the binary classification detection of breathing presence in the environment across various detection time intervals. As elaborated in Fig. \ref{fig:det_time}, the analysis reveals that when the detection time exceeds 7 seconds, accurate identification of potential breathing signals within the detection environment can be achieved with the accuracy of 100\%. Also, the experiment verify the robustness of the proposed approach by changing the the distance between the target and the closest receive. Note that analytical approximation is achieved by logistic fitting curve, which can better reveal the tendency of real accuracy. The definition of the accuracy is shown in (\ref{accuracy}): 
\begin{equation}\label{accuracy}
    Accuracy = \frac{TN+TP}{All\ Predicted\ Labels},
\end{equation}
where $TN$ and $TP$ denote the number of correctly classified positive samples, and negative samples, respectively. Moreover, the classification accuracy increases with respect to the length of detection time, in line with the expectations. 
\begin{table*}[ht]
  \centering
  \caption{Comparison of Studies}
  \label{table2}
  \begin{tabular}{lccccc}
    \toprule
    \textbf{Name} & \textbf{Working frequency} & \textbf{System} & \textbf{Interference}& \textbf{Accuracy} & \textbf{Application}  \\
    \midrule
    Zhang et al.\cite{zengFullBreatheFullHuman2018} & WiFi & 2 commercial WiFi & No & - &Human Respiration\\
    MIT CSAIL\cite{zhao_through-wall_2018} & Milimeter Wave & 2 rows of FMCW radar & No &96\% & Human skeleton\\
    Sun et al.\cite{sun_through-wall_2021} & WiFi & Commercial WiFi and Horn Antenna & No & - & Human Motion\\
    Ali et al.\cite{ali_early_2020} & Milimeter Wave & Simulation & No & - & Vehicle Motion\\
    \textbf{Ours} & \textbf{Milimeter} & \textbf{4 Sivers phased array} & \textbf{Yes} & \textbf{90\%} & \textbf{Human Respiration}\\
    \bottomrule
  \end{tabular}
\end{table*}
\subsubsection{Counting}
\begin{figure}[ht] 
	\centering  
	\vspace{-0.35cm} 
	\subfigtopskip=2pt 
	\subfigbottomskip=2pt 
	\subfigcapskip=-5pt 
	\subfigure[]{
		\label{3}
		\includegraphics[width=0.48\linewidth]{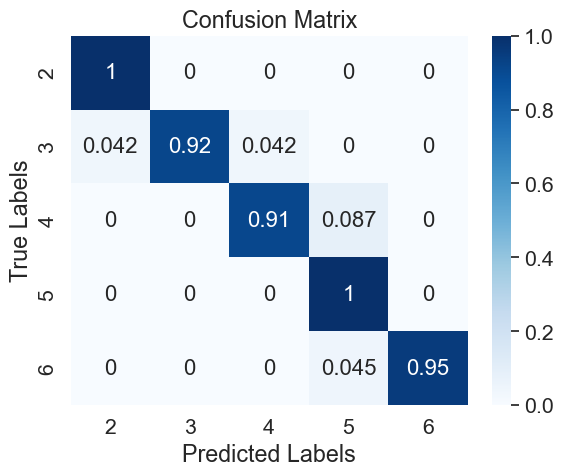}}
    \hspace*{-2mm}
    \subfigure[]{
		\label{7}
		\includegraphics[width=0.48\linewidth]{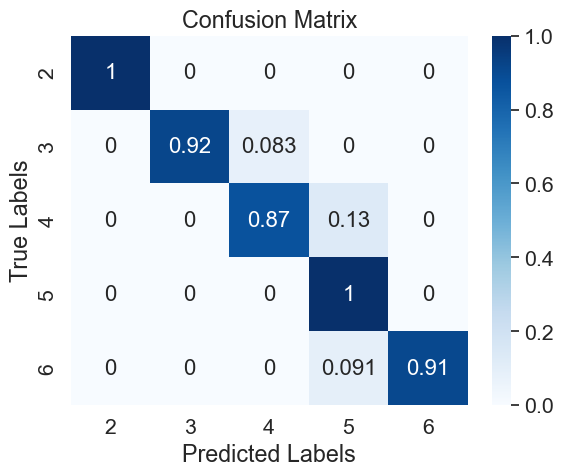}}
    \hspace*{-2mm}
    	\subfigure[]{
		\label{star}
		\includegraphics[width=0.48\linewidth]{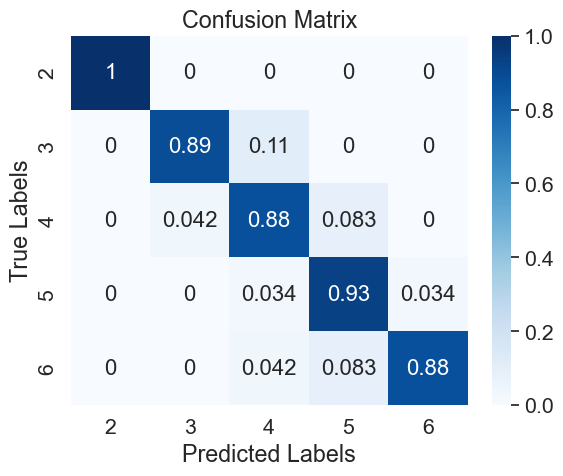}}
    \hspace*{-2mm}
	\subfigure[]{
		\label{star}
		\includegraphics[width=0.48\linewidth]{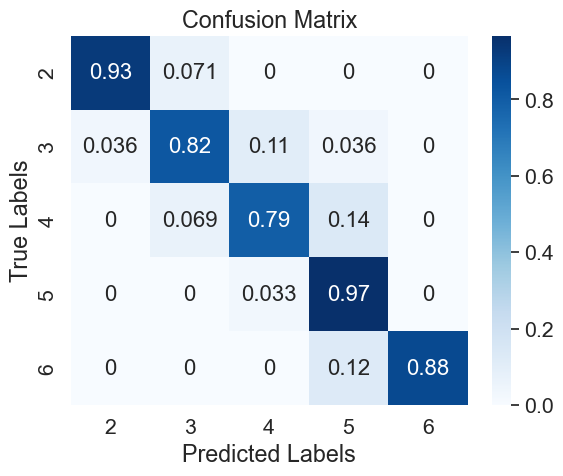}}
	\caption{Confusion matrix for counting results (Although it is a regression task, the confusion matrix is used to evaluate the accuracy of the rounded regression results). (a) Represents the results of the distance between the target and the closest receiver is 0.3 m. The distance in (b), (c) and (d) are 0.6 m, 0.9 m and 1.2 m respectively.}

	\label{fig:resnet_output}
\end{figure}

In the task of counting the human respiration, five experimental scenarios are performed: 2 respirations, 3 respirations, 4 respirations, 5 respirations and 6 respirations within 10 seconds. In order to achieve this counting task, we train the network with 800 samples (160 samples per type) and treat remaining samples as the test set. The results with the closest distance is 0.3 m are illustrated in Fig. \ref{fig:resnet_output}(a). The experimental results show that the accuracy of counting the human respiration rate achieve around 90\%.


The comparison of other studies is shown in Table \ref{table2}.

\section{Conclusion}  
\label{Conclusion}
This paper presents a comprehensive study on an ISAC system operating in the 60.48 GHz frequency band. The performance of the system in counting human respiration under interference is demonstrated by executing five distinct respiratory rates in the surveillance channel and collecting a dataset of received signals using the aforementioned system. Subsequently, a ResNet model is trained using the collected dataset for both respiration detection and respiratory rate counting. The experimental results showcase remarkable classification accuracy, approaching 100\%, for detection times exceeding 6 seconds. The accuracy of counting the frequency of human respiration achieve 90\% overall.

\scriptsize
\bibliographystyle{IEEEtran}

\end{document}